\documentclass[showpacs,preprintnumbers,pre,a4paper,twocolumn]{revtex4}
\usepackage{amssymb}
\usepackage{amsfonts}
\usepackage{amsmath}
\usepackage{graphicx}
\usepackage{dcolumn}
\usepackage{bm}

\input{tcilatex}

\begin{document}

\preprint{APS/123-QED}
\title{Optimal phase space projection for noise reduction}
\author{Xiaodong Luo}
\email{enxdluo@eie.polyu.edu.hk}
\author{Jie Zhang}
\author{Michael Small}
\affiliation{Department of Electronic and Information Engineering, Hong Kong Polytechnic
University, Hom Hung, Hong Kong.}
\date{\today }

\begin{abstract}
In this communication we will re-examine the widely studied technique of
phase space projection. By imposing a time domain constraint (TDC) on the
residual noise, we deduce a more general version of the optimal projector,
which includes those appearing in previous literature as subcases but does
not assume the independence between the clean signal and the noise. As an
application, we will apply this technique for noise reduction. Numerical
results show that our algorithm has succeeded in augmenting the
signal-to-noise ratio (SNR) for simulated data from the 
R\"ossler
system and experimental speech record.
\end{abstract}

\maketitle

\section{Introduction}

Due to its simplicity in implementation and efficiency in computation, noise
reduction based on phase space projection has been widely studied in
previous literature. For example, Broomhead and King \cite{broomhead
extracting} advocated that, in case of white noise, via singular value
decomposition (SVD), one could extract qualitative dynamics from
experimental (noisy) time series by removing the empirical orthogonal
functions (EOFs) \cite{vautard singular} of the trajectory matrix which
correspond to the noise components. To deal with the case of colored noise,
Allen and Smith \cite{allen optimal} proposed a more general method, which
would statistically pre-whiten colored noise by introducing a transformation
to the covariance matrix of noise. In general, phase space projection based
on these methods would not operate on the EOFs that span the
signal-plus-noise subspace, therefore those operations could achieve a
lowest possible distortion for the clean signal, but at the price of a
highest possible residual noise level \cite{ephraim signal}. To obtain an
optimal tradeoff between signal distortion and residual noise so as to
minimize the overall distortion, Ephraim and Trees proposed the time domain
constraint (TDC) projector \cite{ephraim signal}, which improves the
performance of the existing methods by imposing a constraint on the residual
noise, and which also includes the existing methods as its subcases. As a
generalization, some authors also extended the TDC projector to the cases
with colored noise 
\cite{doclo multimicrophone,hu generalized}%
.

Usually, these authors will make two assumptions concerning the experimental
time series. The first assumption is that the time series is stationary and
ergodic, and the second one is that the noise components are independent of
the clean signal. In this communication we will re-examine the idea of the
TDC projector and deduce a more universal version. We will also show that,
with the first assumption, the second is not necessary in general.

The remainder of this article will go as follows: In the second section we
will introduce the idea of the TDC projector. Based on the assumption that
the noisy time series is stationary and ergodic, we will obtain the optimal
TDC projector for a trajectory matrix in the sense of minimizing signal
distortion subject to a permissible noise level. In the third section we
will apply the optimal TDC projector to simulated data from the R\"{o}ssler
system and experimental speech data. We will also compare the performance of
the projectors under different TDCs. Finally, a conclusion is available to
summarize the whole article.

\section{Mathematical deduction}

Given a noisy time series $s=\{s_{i}\}_{i=1}^{M}$, we suppose that the
corresponding clean signal and the additive noise components are $%
d=\{d_{i}\}_{i=1}^{M}$ and $n=\{n_{i}\}_{i=1}^{M}$ respectively, thus for
each noisy data point $s_{i}$, we have $s_{i}=d_{i}+n_{i}$. In addition, we
assume $\{s_{i}\}_{i=1}^{M}$ are (weakly) stationary and ergodic so that its
expectation exists and its variance is finite, while its (auto)covariances
only depend on the time difference between the subsets.

Following the definition in \cite{broomhead extracting}, we could construct
a $(M-m+1)\times m$ trajectory matrix $\mathbf{S}$ from $\{s_{i}\}_{i=1}^{M}$
by letting 
\begin{equation*}
\mathbf{S}=\left( 
\begin{array}{cccc}
s_{1} & s_{2} & ... & s_{m} \\ 
s_{2} & s_{3} & ... & s_{m+1} \\ 
\vdots &  & \ddots &  \\ 
s_{M-m+1} & s_{M-m+2} & ... & s_{M}%
\end{array}%
\right) _{(M-m+1)\times m}
\end{equation*}%
with $M-m+1>m$. Similarly, we could also obtain the corresponding trajectory
matrices $\mathbf{D}$ and $\mathbf{N}$ for components $\{d_{i}\}_{i=1}^{M}$
and $\{n_{i}\}_{i=1}^{M}$ respectively, and we have $\mathbf{S}=\mathbf{D}+%
\mathbf{N}$.

For the purpose of noise reduction, we introduce a projection operator $%
\mathbf{H}$ on the trajectory matrix $\mathbf{S}$ of noisy signal, through
which we could obtain a matrix $\mathbf{Z=SH}$. We define $\mathbf{R}_{0}=$ $%
\mathbf{Z-D=D(H-I}_{m}\mathbf{)}+\mathbf{NH}$ as the matrix of residual
signal, where the term $\mathbf{D(H-I}_{m}\mathbf{)}$ means signal
distortion and the term $\mathbf{NH}$ is residual noise. With the intention
of data augmentation, we would require to achieve as small signal distortion
as possible. Thus $\mathbf{H}=\mathbf{I}_{m}$ would be an intuitive choice.
However, in situations such as speech communication, one would also require
a permissible residual noise level of the noisy signal, and the objective
becomes to minimize signal distortion subject to achieving a permissible
residual noise level. Thus if the initial data does not fulfil this
requirement, one has to reduce the initial noise level at the price of
introducing possible signal distortion. Similar to the idea proposed in \cite%
{ephraim signal}, here we impose a time domain constraint (TDC) $\mu $ on
the term of residual noise $\mathbf{NH}$ and treat $\mathbf{R=D(H-I}_{m}%
\mathbf{)}+\mu \mathbf{NH}$ as the part that requires a minimal distortion,
where $\mu ^{2}\in \lbrack 0,+\infty )$ is the Lagrange multiplier
determined by the permissible noise level from the practical demand (see Eq.
(33) of \cite{ephraim signal} and the related discussions therein). Thus our
objective will be to minimize the average energy $\Xi =\left(
\sum\limits_{i=1}^{M}r_{i}^{2}\right) \left/ M\right. $ of the data set $%
r=\{r_{i}\}_{i=1}^{M}$ that (approximately) corresponds to the matrix $%
\mathbf{R}$. If $M\gg $ $m$, then

\begin{equation}
\Xi \approx \frac{1}{(M-m+1)m}tr(\mathbf{R}^{T}\mathbf{R})\text{,}
\label{residual_signal_energy}
\end{equation}%
where $tr(\cdot )$ means the trace of a square matrix, $\mathbf{R}^{T}$
denotes the transpose of the matrix $\mathbf{R}$.

Discarding the constant term $tr(\mathbf{D}^{T}\mathbf{D})$ in $tr(\mathbf{R}%
^{T}\mathbf{R})$, we have%
\begin{eqnarray}
tr(\mathbf{R}^{T}\mathbf{R}) &=&tr(\mathbf{H}^{T}(\mathbf{D}+\mu \mathbf{N}%
)^{T}(\mathbf{D}+\mu \mathbf{N})\mathbf{H})  \notag \\
&&-2tr(\mathbf{H}^{T}(\mathbf{D}+\mu \mathbf{N})^{T}\mathbf{D}).
\end{eqnarray}%
Taking $m$ as a constant \cite{note0}, for the minimization problem, by
requiring $\partial tr(\mathbf{R}^{T}\mathbf{R})/\partial \mathbf{H}=\mathbf{%
0}$, we would have $(\mathbf{D}+\mu \mathbf{N})^{T}(\mathbf{D}+\mu \mathbf{N}%
)\mathbf{H-}(\mathbf{D}+\mu \mathbf{N})^{T}\mathbf{D=0}$ according to the
differential rules in, for example, 
\cite[p. 472]{Lutkepohl introduction}%
. Therefore the optimal projector 
\begin{equation}
\mathbf{H}_{\min }=\left\{ (\mathbf{D}+\mu \mathbf{N})^{T}(\mathbf{D}+\mu 
\mathbf{N})\right\} ^{-1}(\mathbf{D}+\mu \mathbf{N})^{T}\mathbf{D}\text{.}
\label{H_min}
\end{equation}%
With the noise components, $\partial tr(\mathbf{R}^{T}\mathbf{R})/\partial 
\mathbf{H}^{2}=2(\mathbf{D}+\mu \mathbf{N})^{T}(\mathbf{D}+\mu \mathbf{N})$
is positive definite, which confirms that the extremum taken at $\mathbf{H}%
_{\min }$ is a minimum. The corresponding minimal value%
\begin{equation}
tr_{\min }(\mathbf{R}^{T}\mathbf{R})=tr(\mathbf{D}^{T}\mathbf{D})-tr(\mathbf{%
D}^{T}(\mathbf{D}+\mu \mathbf{N})\mathbf{H}_{\min }\mathbf{)}\text{.}
\label{tr_min}
\end{equation}%
But note that $(\mathbf{D}+\mu \mathbf{N})^{T}$ is not a square matrix, its
(ordinary) inverse matrix usually is not defined, thus we could not cancel
the terms of $(\mathbf{D}+\mu \mathbf{N})^{T}$ in Eq. (\ref{H_min}).

Since $\mathbf{S=D}+\mathbf{N}$, we could also write Eq. (\ref{H_min}) in
the form of\ \ \ \ \ \ \ \ \ \ \ \ \ \ \ \ \ \ \ \ \ \ \ \ \ \ \ \ \ \ \ \ \
\ \ \ \ \ \ \ \ \ \ \ \ \ \ \ \ \ \ \ \ \ \ \ \ \ \ \ \ \ \ \ \ \ \ \ \ \ \
\ \ \ \ \ \ \ \ \ \ \ \ \ \ \ \ \ \ \ \ \ \ \ \ \ \ \ \ \ \ \ \ \ \ \ \ \ \
\ \ \ \ \ \ \ \ \ \ \ \ \ \ \ \ \ \ \ \ \ \ \ \ \ \ \ \ \ \ \ \ \ \ \ \ \ \
\ \ \ \ \ \ \ \ \ \ \ \ \ \ \ \ \ \ \ \ \ \ \ \ \ \ \ \ \ \ \ \ \ \ \ \ \ \
\ \ \ \ \ \ \ \ \ \ \ \ \ \ \ \ \ \ \ \ \ \ \ \ \ \ \ \ \ \ \ \ \ \ \ \ \ \
\ \ \ \ \ \ \ \ \ \ \ \ \ \ \ \ \ \ \ \ \ \ \ \ \ \ \ \ \ \ \ \ \ \ \ \ \ \
\ \ \ \ \ \ \ \ \ \ \ \ \ \ \ \ \ \ \ \ \ \ \ \ \ \ \ \ \ \ \ \ \ \ \ \ \ \
\ \ \ \ \ \ \ \ \ \ \ \ \ \ \ \ \ \ \ \ \ \ \ \ \ \ \ \ \ \ \ \ \ \ \ \ \ \
\ \ \ \ \ \ \ \ \ \ \ \ \ \ \ \ \ \ \ \ \ \ \ \ \ \ \ \ \ \ \ \ \ \ \ \ \ \
\ \ \ \ \ \ \ \ \ \ \ \ \ \ \ \ \ \ \ \ \ \ \ \ \ \ \ \ \ \ \ \ \ \ \ \ \ \
\ \ \ 
\begin{eqnarray}
\mathbf{H}_{\min } &=&\left\{ (\mathbf{S}+(\mu -1)\mathbf{N})^{T}(\mathbf{S}%
+(\mu -1)\mathbf{N})\right\} ^{-1}  \notag \\
&&\times (\mathbf{S}+(\mu -1)\mathbf{N})^{T}\mathbf{(\mathbf{S}-\mathbf{N})}%
\text{.}  \label{H_min_inSN}
\end{eqnarray}

If we assume the clean signal and the noise components are independent,
statistically we have $\mathbf{D}^{T}\mathbf{N=N}^{T}\mathbf{D=0}$ as $%
M\rightarrow \infty $, hence $\mathbf{S}^{T}\mathbf{S=D}^{T}\mathbf{D+N}^{T}%
\mathbf{N}$, and Eq. (\ref{H_min_inSN}) reduces to%
\begin{equation}
\mathbf{H}_{\min }=\left\{ \mathbf{S}^{T}\mathbf{S+(}\mu ^{2}-1\mathbf{)N}%
^{T}\mathbf{N}\right\} ^{-1}(\mathbf{S}^{T}\mathbf{S-\mathbf{N}^{T}\mathbf{N}%
)}\text{.}  \label{H_min_uncorrelated_SN}
\end{equation}

Let $\mathbf{C}_{S}$, $\mathbf{C}_{N}$ denote the covariance matrices of $%
\{s_{i}\}_{i=1}^{M}$ and $\{n_{i}\}_{i=1}^{M}$ respectively, by assuming the
expectation values $E(s)=E(n)=0$, we have $\mathbf{C}_{S}=\mathbf{S}^{T}%
\mathbf{S/(}M-m+1\mathbf{)}$ and $\mathbf{C}_{N}=\mathbf{N}^{T}\mathbf{N/(}%
M-m+1\mathbf{)}$ as $M\rightarrow \infty $. Thus Eq. (\ref%
{H_min_uncorrelated_SN}) would be expressed as

\begin{equation}
\mathbf{H}_{\min }=\left\{ \mathbf{C}_{S}+\mathbf{(}\mu ^{2}-1\mathbf{)C}%
_{N}\right\} ^{-1}(\mathbf{C}_{S}\mathbf{-\mathbf{C}_{N})}\text{,}
\label{H_min_uncorrelated_C}
\end{equation}%
which is consistent with the result in, for example, Eq. (3) of \cite{hu
generalized}. But note that here we use $\mu ^{2}$ to substitute for the
multiplier $\mu $ in Eq. (3) of 
\cite{hu generalized}%
. Also note that $\mathbf{H}_{\min }$ in our work is the transpose of that
in Eq. (3) of 
\cite{hu generalized}%
, this is because the trajectory matrices in our work are essentially the
transpose of those in 
\cite{doclo multimicrophone,ephraim signal,hu generalized}%
.

In many situations, although the noise components are theoretically
uncorrelated to the clean signal, numerical calculations often indicate that
the assumption $\mathbf{D}^{T}\mathbf{N=N}^{T}\mathbf{D=0}$ does not hold
strictly for finite data sets. As a more rigorous form, Eq. (\ref{H_min_inSN}%
) needs no independence assumption between the noise components and the
clean signal. Thus this expression is a further generalization of previous
studies.

\section{Numerical results}

We note that the trajectory matrices previously introduced are all Hankel
matrices. Take trajectory matrix $\mathbf{S}$ of the noisy signal as an
example, its entries satisfy $\mathbf{S}(i,j)=\mathbf{S}(k,l)$ if $i+j=k+l$,
where $\mathbf{S}(i,j)$ denote the element of matrix $\mathbf{S}$ on $i$-th
row and $j$-th column. However, matrix $\mathbf{Z=SH}$ usually will not be a
Hankel matrix, and we may have many ways to obtain the filtered (or
projected) signal $\{z_{i}\}_{i=1}^{M}$. In our work we use the method of
secondary diagonal averaging to extract signal from the matrix $\mathbf{Z}$,
which takes the average of the elements along the secondary diagonals of
matrix $\mathbf{Z}$ as the filtered signal $\{z_{i}\}_{i=1}^{M}$ (for
details, see 
\cite[p. 24]{golyandina analysis}%
), and thus can form a new trajectory (Hankel) matrix $\mathbf{Z}^{H}$ from $%
\{z_{i}\}_{i=1}^{M}$. Golyandina \textit{et al}. prove that this method is
optimal among all Hankelization procedures in the sense that the matrix
difference $\mathbf{Z}^{H}-\mathbf{Z}$ has minimal Frobenius norm 
\cite[p. 24, p. 266]{golyandina analysis}%
.

\begin{table*}[t]
\centering
\parbox{4.2in}{\caption{\label{table1}Performance of TDC projectors for the R\"ossler system
(in unit of dB). }}
\begin{tabular*}{4.2in}{cccc}
\hline\hline
TDC $\mu $ & Additive white noise & Additive colored noise & Multiplicative
noise \\ \hline
& \multicolumn{1}{l}{$20\rightarrow 25.50\pm 0.09$} & \multicolumn{1}{l}{$20\rightarrow 20.92\pm 0.05$} & \multicolumn{1}{l}{$20\rightarrow 21.11\pm
0.10$} \\
0.0 & \multicolumn{1}{l}{$10\rightarrow 15.95\pm 0.11$} & \multicolumn{1}{l}{$10\rightarrow 10.89\pm 0.04$} & \multicolumn{1}{l}{$10\rightarrow 11.14\pm
0.09$} \\
& \multicolumn{1}{l}{$0\rightarrow 5.88\pm 0.06$} & \multicolumn{1}{l}{$0\rightarrow 0.87\pm 0.04$} & \multicolumn{1}{l}{$0\rightarrow 1.16\pm 0.10$}
\\
& \multicolumn{1}{l}{} & \multicolumn{1}{l}{} & \multicolumn{1}{l}{} \\
& \multicolumn{1}{l}{$20\rightarrow 25.80\pm 0.10$} & \multicolumn{1}{l}{$20\rightarrow 21.07\pm 0.05$} & \multicolumn{1}{l}{$20\rightarrow 23.23\pm
0.24$} \\
0.5 & \multicolumn{1}{l}{$10\rightarrow 17.74\pm 0.15$} & \multicolumn{1}{l}{$10\rightarrow 11.64\pm 0.06$} & \multicolumn{1}{l}{$10\rightarrow 14.42\pm
0.23$} \\
& \multicolumn{1}{l}{$0\rightarrow 9.71\pm 0.10$} & \multicolumn{1}{l}{$0\rightarrow 3.12\pm 0.08$} & \multicolumn{1}{l}{$0\rightarrow 6.97\pm 0.22$}
\\
& \multicolumn{1}{l}{} & \multicolumn{1}{l}{} & \multicolumn{1}{l}{} \\
& \multicolumn{1}{l}{$20\rightarrow 26.27\pm 0.11$} & \multicolumn{1}{l}{$20\rightarrow 21.17\pm 0.05$} & \multicolumn{1}{l}{$20\rightarrow 24.44\pm
0.34$} \\
1.0 & \multicolumn{1}{l}{$10\rightarrow 18.29\pm 0.16$} & \multicolumn{1}{l}{$10\rightarrow 11.89\pm 0.07$} & \multicolumn{1}{l}{$10\rightarrow 16.12\pm
0.32$} \\
& \multicolumn{1}{l}{$0\rightarrow 10.10\pm 0.09$} & \multicolumn{1}{l}{$0\rightarrow 4.15\pm 0.09$} & \multicolumn{1}{l}{$0\rightarrow 9.56\pm 0.33$}
\\ \hline\hline
\end{tabular*}
\end{table*}%

We adopt the signal-to-noise ratio (SNR) as the metric to evaluate the
performance of our noise reduction scheme, which is defined (in dB) as 
\cite{ephraim signal,johnson generalized}%

\begin{equation}
SNR=10\log _{10}\frac{\left\Vert d\right\Vert ^{2}}{\left\Vert
z-d\right\Vert ^{2}}\text{,}  \label{SNR_1}
\end{equation}%
where $\left\Vert d\right\Vert ^{2}=\sum\limits_{i=1}^{M}d_{i}^{2}$ and $%
\left\Vert z-d\right\Vert ^{2}=$ $\sum\limits_{i=1}^{M}(z_{i}-d_{i})^{2}$.

We first apply our algorithm to a simulated data set, which is generated
from the $x$ component of the R\"{o}ssler system

\begin{equation}
\left\{ 
\begin{array}{l}
\dot{x}=-(y+z) \\ 
\dot{y}=x+ay \\ 
\dot{z}=b+(x-c)z%
\end{array}%
\right.   \label{rossler}
\end{equation}%
with parameter $a=0.15$, $b=0.2$ and $c=10$. The data is evenly sampled for
every $0.1$ time units. We generate $10,000$ data points and discard the
first 1000 to avoid transition. To construct the trajectory matrices, we
will set the window size $m=20$ .\ 
\begin{figure*}[t]
\centering
\includegraphics[width=5in]{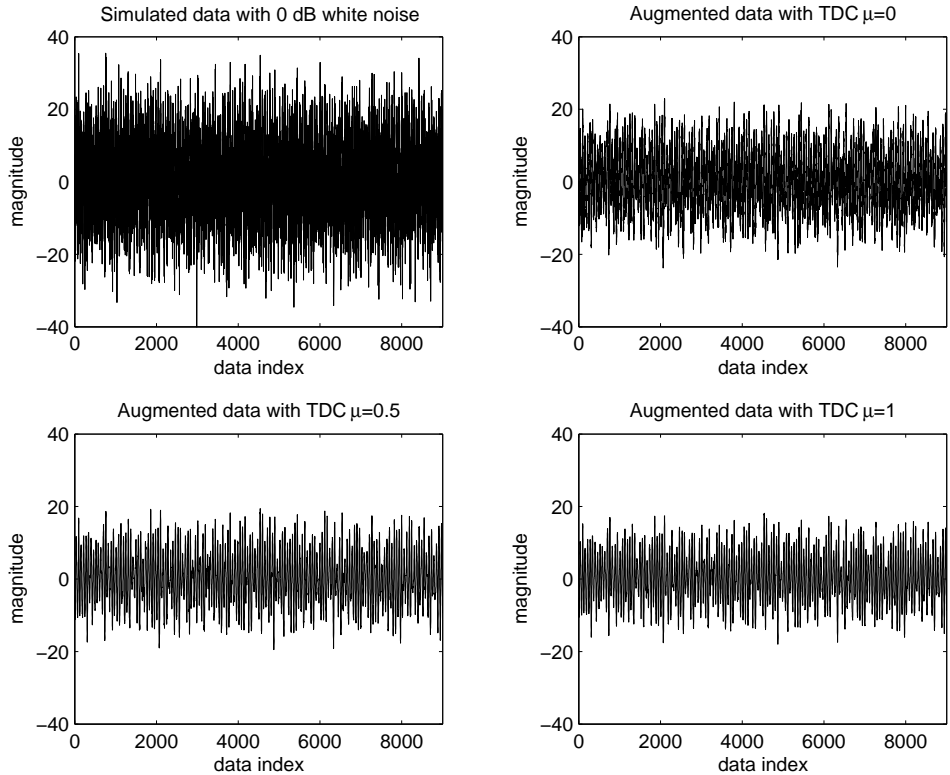}
\caption{(a)  Time series from the R\"{o}ssler system contaminated with 0 dB additive
white noise; (b), (c) and (d)
Augmented time series by TDC projectors with $\mu=
0, 0.5$ and $1$ separately.}
\label{rosslerfig}
\end{figure*}%

Let $\left\{ s_{i}\right\} _{i=1}^{M}$ and $\left\{ d_{i}\right\} _{i=1}^{M}$
again denote the noisy and clean signals respectively. We consider adding
three types of noise contamination to the clean data. The first one is
additive white noise $\left\{ \xi _{i}\right\} _{i=1}^{M}$ (so that $%
s_{i}=d_{i}+\xi _{i}$), which follows the normal Gaussian distribution $%
N(0,1)$. The second one is additive colored noise $\left\{ \eta _{i}\right\}
_{i=1}^{M}$ (so that $s_{i}=d_{i}+\eta _{i}$), which, as an example, is
produced from a third order autoregressive ($AR(3)$) process in the form of $%
\eta _{i}=0.8\eta _{i-1}-0.5\eta _{i-2}+0.6\eta _{i-3}+\xi _{i}$, where
variable $\xi $ follows the normal distribution $N(0,1)$. The last one is
multiplicative noise $\left\{ \zeta _{i}d_{i}\right\} _{i=1}^{M}$ (so that $%
s_{i}=(1+\zeta _{i})d_{i}$). As an example, we let $\zeta _{i}=\eta _{i}^{2}$%
, where $\left\{ \eta _{i}\right\} _{i=1}^{M}$ is from the previous $AR(3)$
process, then the noise component $\left\{ \zeta _{i}d_{i}\right\}
_{i=1}^{M} $ is correlated to the clean data $\left\{ d_{i}\right\}
_{i=1}^{M}$.

By varying the magnitude of the introduced noise, we have the initial noise
level be $20$ dB, $10$ dB, $0$ dB respectively, and for each noise level, we
will include $10$ different noise samples from the same process in
calculation. We will also study the performance of the projectors under
different constraints. As examples, we let TDC $\mu =0$, $0.5$ and $1$
separately. TDC $\mu =0$ will lead to the least-squares (LS) projector based
on the SVD technique that appeared in, for example, 
\cite{allen optimal,broomhead extracting,vautard singular}%
. We would need to specify the dimension of the signal-plus-noise subspace
so as to group the EOFs and eigenvalues that correspond to the noisy signal
and remove the complementary noise subspace, which is essentially related to
the problem of choosing the embedding dimension for embedding reconstruction
from a scalar time series (see the discussion in \cite{johnson generalized}%
). Thus here we adopt the criterion of false nearest neighbor \cite{Kennel},
a method proposed for selection of appropriate embedding dimensions. To
apply this criterion in calculation, we utilized the codes implemented in
the TISEAN package \cite{Tisean} and found that the proper dimension size $K$
of the signal-plus-noise subspace is $5$ in our cases. For $\mu =1$, we will
obtain the well-know linear minimum mean-squared-error (LMMSE) projector
(detailed introductions available in, e.g., \cite{trees detection}). After
all of the calculations, we finally list the performance of these TDC
projectors in Table \ref{table1}. For better comprehension of the presented
results, we provide the waveforms of all of the data listed in Table \ref%
{table1} as the supplementary material \cite{supplementary}. To keep our
presentation concise, here we only take out the raw data contaminated with $0
$ dB additive white noise as an example and depict its waveform of in panel $%
(a)$ of Fig. (\ref{rosslerfig}). For comparison, we also plot the augmented
data with TDC$=0,0.5$ and $1$ in panel $(b)$, $(c)$ and $(d)$, whose mean
noise levels are $5.88,$ $9.71$ and $10.10$ dB correspondingly.

\begin{figure*}[t]
\centering
\includegraphics[width=5in]{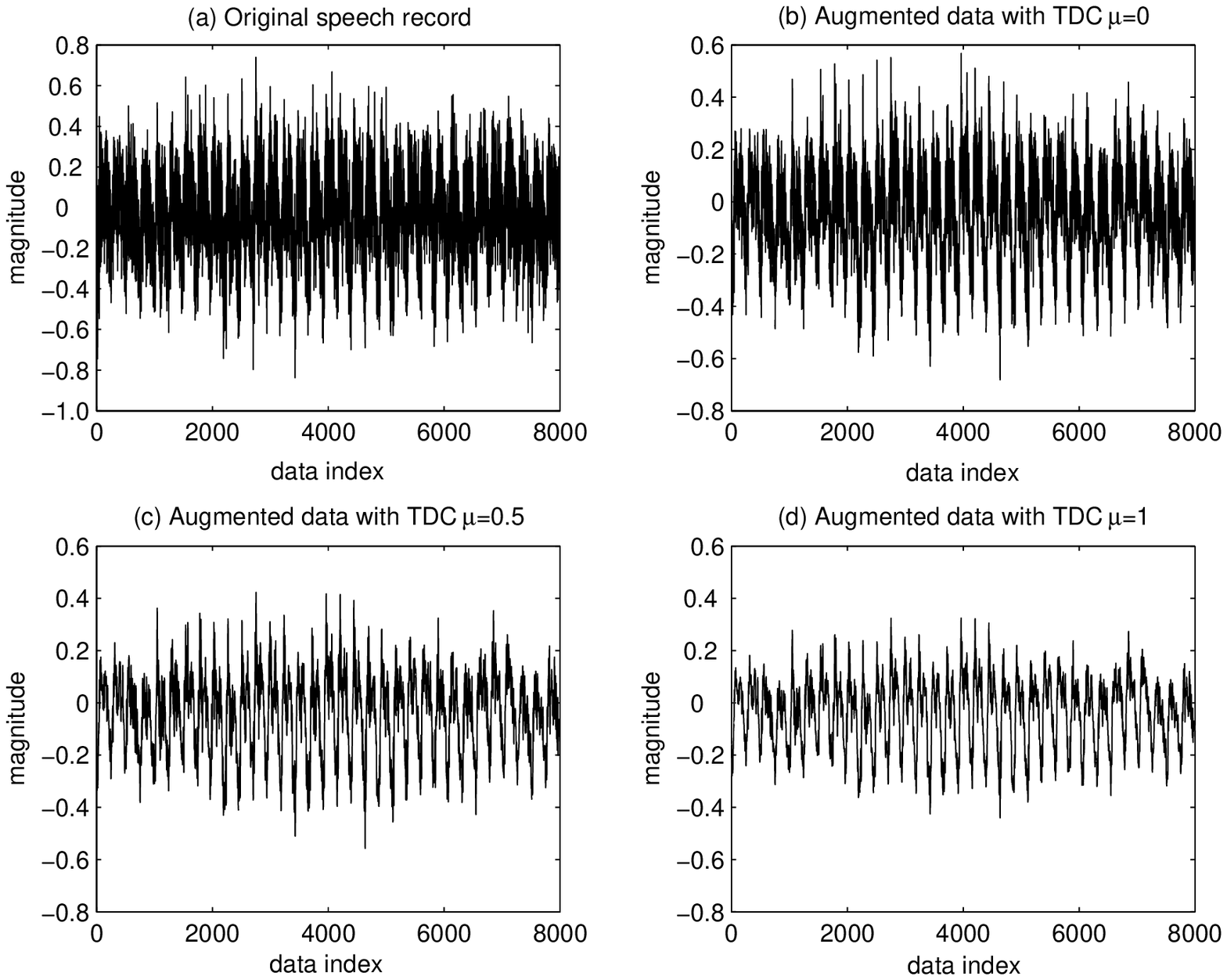}
\caption{(a) Original speech record; (b), (c) and (d)
Speech data output from TDC projectors with $\mu=
0, 0.5$ and $1$ separately.}
\label{augmented_speech}
\end{figure*}%

From Table \ref{table1}, we see that for the R\"{o}ssler system, our
algorithm works for all of the three types of contamination. But the data
augmentation for additive colored noise is not as obvious as those for
additive white noise and multiplicative noise (the possible explanation is
explored in the appendix). We also see that, in general, the LMMSE projector
has better performance than that of the LS projector in the sense that it
can achieve better SNR as defined in Eq. (\ref{SNR_1}).

We then apply our algorithm to a very noisy speech (vowel) data (with $8,000$
data points), which is sampled at $44$ kHz and quantized to $16$ bits. In
this case we only know the background noise measured in the period without
the signal. It would be preferred if we could produce a set of samples that
mimic the behavior of the underlying noise. Here we adopt the
pseudo-periodic surrogate (PPS) algorithm \cite{small surrogate} to generate 
$9$ surrogates based on the original background noise. With these data sets,
the initial SNR of the speech data is estimated to be $-0.32\pm 0.18$ dB via
Eq.(\ref{SNR_1}). To introduce phase space projection to the speech data, we
let the window size $m=30$ and set the dimension size of signal-plus-noise
subspace to be $K=8$, and then apply the TDC projectors $\mathbf{H}$ to its
trajectory \ matrix. For the LS projector ($\mu =0$), the augmented SNR$%
=4.36\pm 0.41$ dB. While for TDC $\mu =0.5$ and $1$, the corresponding SNRs
increase to $6.28\pm 0.61$ dB and $6.97\pm 0.66$ dB respectively. As an
illustration, we plot the waveforms of the original speech record and three
projected data under different TDCs in Fig. (\ref{augmented_speech}), from
which we can see that, the LMMSE projector ($\mu =1$) would lead to a
smoother speech waveform (panel (d)) than that of the LS projector (panel
(b)). Although the speech data output from the LMMSE projector has lower
(signal) magnitudes than those of the speech record from the LS projector,
it is still preferred to its rival in speech communication since a smoother
data will usually bring better communication quality.

\section{Conclusion}

In this communication we re-examined the noise reduction technique based on
phase space projection. By imposing a constraint on the residual noise, we
deduced the optimal time domain constrained projector in the sense of
minimizing signal distortion subject to a permissible noise level. We also
showed that, in general we need not assume independence between clean signal
and noise components as was previously done. This viewpoint was confirmed by
our numerical results (see the third column of the calculation results in
Table \ref{table1}).

\section*{Appendix}

Here let us examine the metric of signal-to-noise ratio (SNR) in more
detail. According to the definition in Eq. (\ref{SNR_1}), $SNR=10\log
_{10}\left\Vert d\right\Vert ^{2}/\left\Vert z-d\right\Vert ^{2},$where $%
\left\Vert d\right\Vert ^{2}=\sum\limits_{i=1}^{M}d_{i}^{2}$ and $\left\Vert
z-d\right\Vert ^{2}=$ $\sum\limits_{i=1}^{M}(z_{i}-d_{i})^{2}$. Note that $%
\left\Vert d\right\Vert ^{2}=tr(\mathbf{D}^{T}\mathbf{D})/m$ and $\left\Vert
z-d\right\Vert ^{2}=tr(\mathbf{(Z-D)}^{T}\mathbf{(Z-D))/}m$ as $M\rightarrow
\infty $, thus 
\begin{equation*}
SNR=10\log _{10}tr(\mathbf{D}^{T}\mathbf{D})-10\log _{10}tr(\mathbf{(Z-D)}%
^{T}\mathbf{(Z-D))}.
\end{equation*}

Since $\mathbf{Z=SH}$, we have $tr(\mathbf{(Z-D)}^{T}\mathbf{(Z-D))=}tr(%
\mathbf{H}^{T}\mathbf{S}^{T}\mathbf{SH})-2tr(\mathbf{D}^{T}\mathbf{SH})+tr(%
\mathbf{D}^{T}\mathbf{D})$. For the case that the noise and the clean signal
are independent. substituting the optimal projector $\mathbf{H}_{\min }$
into the expression, it can be shown that $tr_{\min }(\mathbf{(Z-D)}^{T}%
\mathbf{(Z-D))=}tr(\mathbf{D}^{T}\mathbf{D})-tr(\mathbf{H}_{\min }\mathbf{D}%
^{T}\mathbf{D})$. For simplicity, we assume the expectation values $%
E(d)=E(n)=0$, then $\mathbf{C}_{D}=\mathbf{D}^{T}\mathbf{D/(}M-m+1\mathbf{)}$
and $\mathbf{C}_{N}=\mathbf{N}^{T}\mathbf{N/(}M-m+1\mathbf{)}$ as $%
M\rightarrow \infty $, where $\mathbf{C}_{D}$ and $\mathbf{C}_{N}$ are the
covariance matrix of the clean signal and the noise respectively, and $%
\mathbf{H}_{\min }$ can be expressed in the form of Eq. (\ref%
{H_min_uncorrelated_C}), or equivalently, $\mathbf{H}_{\min }=\{\mathbf{C}%
_{D}+\mu ^{2}\mathbf{C}_{N}\}^{-1}\mathbf{C}_{D}$. Therefore in this case,
we have $tr_{\min }(\mathbf{(Z-D)}^{T}\mathbf{(Z-D))=}tr(\mathbf{C}_{D})-tr(%
\mathbf{H}_{\min }\mathbf{C}_{D})$, thus the maximal SNR can be expressed by

\begin{eqnarray}
SNR_{\max } &=&10\log _{10}tr(\mathbf{C}_{D})-10\log _{10}(tr(\mathbf{C}_{D})
\notag \\
&&-tr(\{\mathbf{C}_{D}+\mu ^{2}\mathbf{C}_{N}\}^{-1}\mathbf{C}_{D}^{2})).
\label{SNR_2}
\end{eqnarray}

Through the SVD technique \cite{broomhead extracting}, $\mathbf{C}_{D}$ can
be written as $\mathbf{C}_{D}=\mathbf{V}_{D}\mathbf{\Lambda }_{D}\mathbf{V}%
_{D}^{T}$ , where $\mathbf{V}_{D}$ is the normalized eigenvector matrix of $%
\mathbf{C}_{D}$, and $\mathbf{\Lambda }_{D}$ is a diagonal matrix whose
non-zero elements are the eigenvalues of $\mathbf{C}_{D}$ (in fact $\mathbf{V%
}_{D}^{T}\mathbf{V}_{D}=\mathbf{I}_{m}$ and $\mathbf{C}_{D}\mathbf{V}_{D}=%
\mathbf{V}_{D}\mathbf{\Lambda }_{D}$). Similarly, we have $\mathbf{C}_{N}=%
\mathbf{V}_{N}\mathbf{\Lambda }_{N}\mathbf{V}_{N}^{T}$. Let $\mathbf{V}_{N}=%
\mathbf{V}_{D}\mathbf{P}_{DN}$ (for better comprehension, $\mathbf{P}_{DN}$
can be thought as a kind of projection from $\mathbf{V}_{N}$ on $\mathbf{V}%
_{D}$), then $\mathbf{C}_{N}=\mathbf{V}_{D}\mathbf{P}_{DN}\mathbf{\Lambda }%
_{N}\mathbf{P}_{DN}^{T}\mathbf{V}_{D}^{T}$. Substitute it into Eq. (\ref%
{SNR_2}), we have 
\begin{eqnarray*}
SNR_{\max } &=&10\log _{10}tr(\mathbf{\Lambda }_{D})-10\log _{10}(tr(\mathbf{%
\Lambda }_{D}) \\
&&-tr(\{\mathbf{\Lambda }_{D}+\mu ^{2}\mathbf{P}_{DN}\mathbf{\Lambda }_{N}%
\mathbf{P}_{DN}^{T}\}^{-1}\mathbf{\Lambda }_{D}^{2})).
\end{eqnarray*}

If the noise components are white, we have $\mathbf{\Lambda }_{N}=\sigma ^{2}%
\mathbf{I}_{m}$ (with $\sigma $ being the standard deviation of the noise
process) and $\mathbf{V}_{N}=\mathbf{V}_{D}$ (i.e. $\mathbf{P}_{DN}=\mathbf{I%
}_{m}$) \cite{allen optimal}. However, for the case of colored noise,
usually $\mathbf{P}_{DN}\neq \mathbf{I}_{m}$. Instead it is possible that
the absolute value of the elements in $\mathbf{P}_{DN}$ are relatively
small. Thus even for the same clean signal $\{d_{i}\}_{i}^{M}$, the $%
SNR_{\max }$ performance of the colored noise might be much worse than that
of the white noise. This fact might explain the observation that the results
in Table. I are not that promising for the additive colored noise.

\section*{Acknowledgement}

This research was supported by Hong Kong University Grants Council
Competitive Earmarked Research Grant (CERG) No. PolyU 5216/04E.

\end{document}